\documentclass[preprint,preprintnumbers,amsmath,amssymb]{revtex4}

\usepackage{graphicx}
\usepackage{dcolumn}
\usepackage{bm}

\begin{document}

\preprint{J. Phys. Chem. A  {\bf 110} (2006)}

\title{Theory and Application of Dissociative Electron Capture in Molecular Identification}
\author{Crystal D. Havey$^\dagger$, Mark Eberhart$^\dagger$, Travis Jones$^\dagger$, Kent J. Voorhees$^\dagger$,
James A. Laram\'ee$^*$, Robert B. Cody$^\ddagger$, Dennis P. Clougherty$^{**}$} 
\affiliation{
$^\dagger$ Department of Chemistry \& Geochemistry, Colorado School of Mines, Golden, CO 80401}

\affiliation{$^*$ EAI Corporation, 1308 Continental Drive, Abingdon, MD 21009}
 
 \affiliation{$^\ddagger$ JEOL-USA, Inc., 11 Dearborn Road, Peabody, MA 01960}
 
\affiliation{$^{**}$
Department of Physics, University of Vermont, Burlington, VT 05405-0125}

\date{\today}

\begin{abstract}
The coupling of an electron monochromator (EM) to a mass spectrometer (MS) has created a new analytical technique, EM-MS, for the investigation of electrophilic compounds.  This method provides a powerful tool for molecular identification of compounds contained in complex matrices, such as environmental samples. In particular, EM-MS has been applied to the detection of nitrated aromatic compounds, many of which are potent mutagens and/or carcinogens and are considered environmental hazards.  EM-MS expands the application and selectivity of traditional MS through the inclusion of a new dimension in the space of molecular characteristicsÑthe electron resonance energy spectrum. EM-MS also enhances detection sensitivity as well because the entire electron flux of the proper energy can be delivered into the negative ion resonance that is analytically most useful to solving the problem at hand. However, before this tool can realize its full potential, it will be necessary to create a library of resonance energy scans from standards of the molecules for which EM-MS offers a practical means of detection.  Unfortunately, the number of such standards is very large and not all of the compounds are commercially available, making this library difficult to construct.  Here, an approach supplementing direct measurement with chemical inference and quantum scattering theory is presented to demonstrate the feasibility of directly calculating resonance energy spectra. This approach makes use of the symmetry of the transition-matrix element of the captured electron to discriminate between the spectra of isomers.  As a way of validating this approach, the resonance values for twenty-five nitrated aromatic compounds were measured along with their relative abundance.  Subsequently, the spectra for the isomers of nitrotoluene were shown to be consistent with the symmetry-based model.  The initial success of this treatment suggests that it might be possible to predict negative ion resonances and thus create a library of EM-MS standards.  
\end{abstract}

\maketitle

\section{introduction}

Developing analytical instrumentation with high detection sensitivity and selectivity is crucial to our abilities to solve meaningful analytical problems that challenge the nation. Among the more sensitive analytical techniques are those employing gas chromatography/mass spectrometric analysis (GC/MS), where GC-retention time, mass, and intensity are molecule selective, and therefore useful for molecular identification.  For a number of years, researchers have sought to expand on MS detection methods by using a monochromatic electron beam of varying energy to scan an unknown analyte.  In a process called dissociative electron capture, electrons of specific energies are captured to produce ions of characteristic masses that can be detected as a fragment ion with MS.   In effect, the resonance energy spectrumÑion yield as a function of electron energyÑcreates an additional molecular characteristic to supplement those produced with traditional GC/MS. 

The trochoidal electron monochromator (EM), which uses an axial magnetic field to prevent spreading of the electron beam, was first introduced by Stamatovic and Schultz \cite{Stamatovic1, Stamatovic2}. Illenberger {\it et al.} then modified and interfaced the EM to a mass spectrometer and demonstrated the ability to produce negative ions and ion fragments along with ion yield curves as a function of electron energy using EM-MS \cite{Illenberger1, Illenberger2}.  More recently, Laram\'ee et al. used IllenbergerÕs design to further improve the use of the EM in MS applications \cite{Laramee1, Laramee2}.  This device was reported to allow the ionization energy to be precisely controlled to $\pm$ 0.3 eV with ion currents only slightly below those observed for electron ionization.  

JEOL USA, Inc. (Peabody, MA) has designed and interfaced an EM similar to that of Laram\'ee et al. for use in a commercially available mass spectrometer.  This instrument produces spectra of negative ion current as a function of electron energy (Figure 1).  For example, the three isomers of mono-nitrotoluene show two negative ion resonances at distinct electron energies, viz, 0.5 and 3 eV.  The relative abundance of the lower energy peak is distinct for the 2-nitrotoluene isomer. These spectra suggest that a new analytical approach could be applied for analysis of complex mixtures that are capable of ionization by electron capture and have led analytic chemists to explore the technique more fully \cite{Petrucci1, Petrucci2}.  However, a number of theoretical and technical issues must be addressed before the technique can begin to fill its full potential.  Key among these are: (1) a demonstration that different molecules give rise to different electron resonance spectra (ERS), and (2) that once a spectrum is obtained, it can be associated with the molecule, or class of molecules, from which it arose.  

First steps in addressing issue (1) above have been taken by Laram\'ee et al. \cite{Laramee2} who have used EM-MS to study a limited number of pesticides, explosives, organofluoro compounds, polychlorinated dibenzodioxins, polychlorinated biphenyls, polycyclic aromatic compounds, and nitrated aromatic compounds.  Here, we expand on this work by determining the ERS of twenty-five nitrated aromatic compounds some of which are representative of those found in explosives, diesel soot, and tobacco smoke.  Generally the electron resonance of these compounds is linked to the detachment of a nitro group, NO$_2^-$, which can be identified by scanning the electron energies associated with the m/z 46 peaks \cite{Laramee2, Laramee3, Voorhees1, Voorhees2}.

Addressing issue (2) above is more difficult.  One approach is simply to create a library of ERS for all molecules of potential interest.    Another approach would be the direct calculation of electron resonance spectra that could further serve to supplement experimentally obtained standards. Such an approach is presented and used to distinguish the isomers of nitrotoluene. 

\section{Experimental Methods}
\subsection{Chemicals. } Hexafluorobenzene and all nitroaromatic compounds were purchased from either Sigma-Aldrich (St. Louis, MO) or Accustandard, Inc. (New Haven, CT) and were used without further purification.  The standards were diluted or dissolved into toluene (ChromAR grade, Mallinckrodt Chemicals, Phillipsburg, NJ) at a concentration of 100 ng/$\mu$L.

\subsection{Instrumentation. } A JEOL trochoidal EM has been interfaced to a JEOL MStation JMS 700-T four sector mass spectrometer.  The filament potential, with respect to its center, was scanned from -3 to 12 eV with an electron current of 20 $\mu$A, The energy resolution was $\pm$ 0.4 eV.  A 6.0 kV accelerating voltage was used, the resolution was set to 1000, and the EM source temperature was 280 C.  
The electron energy scale was calibrated using nitrobenzene and hexafluorobenzene standards introduced to the ionization source via a heated reservoir.  The molecular ion of nitrobenzene (m/z 123) was assigned a resonance energy value of 0.06 eV, while the first and second NO$_2^-$ (m/z 46) resonance peaks were assigned values of 1.2 eV and 3.5 eV, respectively \cite{Laramee1}.

For hexafluorobenzene, the molecular ion (m/z 186) was assigned a value of 0.03 eV, while the first and second m/z 167 peaks from the C$_6$H$_5^-$ fragment were assigned 4.5 eV and 8.3 eV, respectively \cite{Laramee4}  One microliter of each sample solution was injected onto an HP 6890 Series GC equipped with an on-column injector port to avoid degradation of the compounds at the inlet.  A 30-m ZB-5 column set to 280 C was used for resonance energy measurements except for 1,3-dinitropyrene, in which the column temperature was set to 300 C.  

The electron energy scans were carried out using a JEOL electron monochromator power supply modified with an energy sweep board.  A Tektronix (Beaverton, OR) TDS2022 two-channel oscilloscope interfaced to the JEOL NIR Energy Sweep Utility program was used for data acquisition.  

\subsection{Calculations. }  All first principle calculations were done using the Amsterdam Density Functional code \cite{Baerends}  The correction to the exchange and correlation specified by Perdew-Wang \cite{Perdew} were used for all calculations.  Triple zeta and double zeta basis sets including polarization terms were used.  The symmetry arguments presented here were not altered by the choice of basis set.  

\subsection{Experimental Results. } 
Table 1 summarizes the major parameters which we believe distinguish one ERS from another:  namely, the energies and relative intensities of both the electron-capture resonances typical of nitrated aromatic groups.  The variation in these parameters, particularly the vanishing of the lower energy resonance for some compounds, e.g. 2-nitrophenol, suggests that these parameters will potentially be useful in molecular identification.  

Unfortunately, as noted before, the extent to which this potential is realized will depend sensitively on the ability to assign a molecule to its spectrum.  In the absence of any clear rules that associate molecular structure with spectra, the enormous task of creating an ERS library must be confronted. Rather than taking on this task in full, an initial attempt has been made here to create a model of dissociative electron capture, from which rules linking structure to ERS might be inferred.  Thus, providing the motivation for future work to directly calculate these spectra.  

\section{A Model for Dissociative Electron Capture }
Begin by considering the reaction
\begin{equation}
 e^-+AB\to A^-+B
\end{equation}
where $AB$ is an isomer of nitrotoluene and  $B$ is the negatively charged nitro fragment and $A$, a toluene radical.  The radical toluene fragment is much more massive than either the electron or the nitro fragment.  Ignore the kinetic energy of the toluene fragment and take as the Hamiltonian 
 \begin{equation}
 H=T_e+T_B+V_{e-A}+V_{AB}+V_{e-B}
 \end{equation}
 where $T_e$ and $T_B$ are the kinetic energies of the electron and the nitro fragment, respectively; $V_{e-A}$ and $V_{e-B}$ are the interactions of the electron with the toluene and the nitro fragments, respectively; and $V_{AB}$ is the interaction of the nitro group and the toluene fragment.
 
 Following the work of Day {\it et al.} \cite{Day}, we identify the toluene fragment to be a massive ``core.'' It is known \cite{Day} that the interactions with the core fragment can be eliminated from the T-matrix element.  In the first order Born approximation, the T-matrix element is given by
\begin{equation}
\langle A^-, B|T|e^-, AB\rangle = \langle \phi, \vec K|V_{e-B}|\vec k, \Phi_b\rangle
\end{equation}
where $|\vec k\rangle$ is the unperturbed plane wave state of the incident electron, $|\phi\rangle$ is the molecular orbital of the core that captures the electron, $|\Phi_b\rangle$ is the bound state of the nitro fragment to the core, and $|\vec K\rangle$ is the plane wave state of the detached nitro fragment.  
 
 With the nitro fragment so deeply bound to the toluene core initially, we make the further approximation that  $\langle\vec R|\Phi_b\rangle\approx\delta(\vec R)$, giving
\begin{equation}
\langle A^-, B|T|e^-, AB\rangle \approx {1\over\sqrt{\Omega}}\langle \phi |V_{e-B}|\vec k\rangle
\end{equation}
 where $\Omega$ is the volume of the box used to normalize the continuum eigenstates.

Lastly, it is recognized, on the basis of ab-initio density functional calculations on nitrotoluene, that the incoming electron is attracted to the phenyl moiety and repelled by the nitro group because a permanent dipole is present in this class of chemical compounds. Such an interaction is long-range, and over the extent of the short-range bound state in which the electron is captured, the electron-nitro interaction can be approximated as a repulsive constant, $V_{e_B}\approx V_0$.

\begin{eqnarray}
\langle A^-, B|T|e^-, AB\rangle &\approx {V_0\over{\Omega}}\int d^3\vec r\  e^{i\vec k\cdot\vec r} \phi^*(\vec r)\cr
&\equiv {V_0\over{\Omega}} \phi^*(\vec k)
\end{eqnarray}

By Fermi's Golden Rule, the reaction rate averaged over all molecular orientations--a quantity closely related to the electron capture resonance spectra-- is given by
\begin{equation}
 \Gamma={V_0^2\over 2\hbar\Omega}\int {d^3\vec K d\Omega_k\over (2\pi)^3} |\phi(\vec k)|^2 \delta(\epsilon_k
 +\epsilon_\phi-E_K-\Delta)
 \end{equation}
 where $\epsilon_k$ is the incident electron kinetic energy, $E_K$ is the final fragment kinetic energy, 
 $\epsilon_\phi$ is the electron binding energy for the capturing orbital, and $\Delta$ is the binding energy of the nitro fragment to the toluene core.  Thus, the essential component needed for the calculation of the resonance spectra from first principles is nothing more than the Fourier transform of the capturing orbital.
 
The difficulty in evaluating this matrix element is ascertaining into which of the many molecular orbitals the incident electron is initially captured.  However, inspection of Table 1 suggests that in the case of nitrated aromatic compounds, attention may be restricted to the unoccupied $\pi$-orbitals.  The rationale behind this conclusion derives from the observation that the resonance energies do not vary substantially among compounds characterized by strikingly different functionalities, e.g. nitrophenol and nitrotoluene, both of which show primary resonances close to 3.6 eV.  Accordingly, the wavefunctions of the capturing molecular orbital are also unlikely to show appreciable variation. 

	The only wavefunctions that could satisfy this constraint are those of the molecular $\pi$-orbitals, which will be characterized by contributions from the $p_z$ orbitals of the nitro group and ring carbons (we take z to be normal to the aromatic ring).  Ring functionalities without $\pi$-electrons, e.g. methyl or hydroxyl groups, will contribute only modestly to these molecular $\pi$-orbitals.  

In the case of substituted phenyl compounds, like the isomers of nitrotoluene, there are three unoccupied $\pi$-orbitals that should be considered, the structure of which can be predicted from simple nodal arguments, Figure 2.  The lowest lying of the three will be the molecular LUMO and possesses two nodal planes, neither of which contains the nucleus of a ring carbon atom.  This forces the nitro-to-ring interaction to be $\pi$-bonding.  The other $\pi$-orbital with two nodal planes is generally the LUMO + 1.  One of its nodal planes contains the ring carbon bound to the nitrogen atom, producing a non-bonding interaction between the nitro group and the ring.  At still higher energies is the fully anti-bonding $\pi$-orbital characterized by three nodal planes.  

By virtue of the placement of its nodal planes, the nitro-to-ring nonbonding orbital will be composed almost entirely of the pz atomic orbitals on the carbon atoms at the 2, 3, 5, and 6 positions relative to the nitro group.  As such, we expect this orbital will be nearly identical in the isomers of nitrotoluene and nitrophenol and thus is likely to be the orbital into which the 3.6 eV electron is captured.

	As a partial confirmation of this conjecture, in Figure 3 isosurfaces are shown for the LUMOs + 1 of 2-, 3-, and 4-nitrotoluene as computed from first principles using the Amsterdam Density Functional code \cite{Baerends}.   As expected, these orbitals are similar and localized on the ring, with very little amplitude on the nitro or methyl functionalities.

With a candidate capturing wavefunction, the energy of an incident electron that will maximize the matrix element  $\langle\vec k|\psi_2\rangle$ must be determined.  Here, $|\psi_2\rangle$ is the state corresponding to the LUMO + 1.  This observation invites the evaluation of the matrix element through a plane wave expansion of the capturing orbital and a direct determination of its overlap with an incident electron.  

	However, as a first step, and as a means of testing our previous assumptions and building intuition, we begin with a semiquantitative analysis, noting that electron capture will be governed by a ``local symmetry'' that may be higher than that of the full molecule.  This local symmetry is that of the potential only on those sites where the wavefunction of the capturing molecular orbital has significant amplitude.  Thus, from Figure 3 it is shown that the symmetry governing the capture of an electron into the LUMO + 1 of 2-, 3-, or 4-nitrotoluene will be that of the potential about the carbon atoms at the 2, 3, 5, and 6 ring positions relative to the nitro group, for it is on these atoms that the LUMO is localized.  Over this region, the symmetry is nearly D$_{2h}$, and it will be this symmetry that governs electron capture into the LUMOs + 1 of the isomers of nitrotoluene.  Under this local symmetry, if the capturing matrix element is to be nonzero, the capturing LUMO must share the same irreducible representation as the wavefunction of the incident electron. 

The LUMOs + 1 of the isomers of nitrotoluene, all transform as the A$_u$ irreducible representation of the D$_{2h}$ point group.  And it is a straightforward exercise to project out that component of a general plane wave with this symmetry.  This exercise reveals the wavefunction of the incident electron to be, $ i \sin(k_xx)\sin(k_yy)\sin(k_z z)$, where $k_x$, $k_y$, and $k_z$ are the Cartesian components of the electronÕs wavevector.  Note that the wavefunction vanishes, if any of these components is zero.

	The Fourier transform of the LUMO + 1 has a substantial magnitude for wavevectors corresponding to a phase shift of $\pi$ along a displacement vector connecting orbital lobes of opposite phase, as depicted in Figure 4A and B.  The magnitude of this wavevector can be estimated by placing the endpoints of the displacement vector in the regions of highest amplitude for the orbital lobes.  This displacement must be a half-wavelength to correspond to having the electron wavefunction undergo a phase change of $\pi$. 

It is now possible to estimate the wavelength of the incident captured electron and hence its energy.  The x-component of the displacement vectors shown in Figure 4A corresponds to the distance between neighboring ring carbon atoms, 1.4 \AA.  The y-component is the distance between carbon atoms meta to each other, 2.42 \AA.  The z-component varies from an upper limit of 2 \AA\ to a lower limit given by the distance between the LUMO extrema, or approximately 1 \AA.  These values yield electron wavelengths between 6.9 and 5.9 \AA, corresponding to wavevectors between, 0.9 and 1.1 \AA$^{-1}$, and electron energies between 3.2 and 4.3 eV respectively.  Using these values for $k_x$, $k_y$, and $k_z$, a double isosurface for the corresponding wavefunction is shown as Figure 4C, which, overlaps constructively over all space with the molecular wavefunction, accounting for the large amplitude of the resonance at 3.6 eV. 

Given the approximations involved, these are remarkable results.  The experimental scans show a broad electron resonance at 3.6 eV, but with a half width of nearly 1.5 eV.  The semiquantitative estimate is well within this margin of error, and correctly predicts the peak resonance energy.  Further, the model attributes the observed width of the spectrum to the broadness of the Fourier transform of the LUMO + 1 around this optimum k-value, which is due, in part, to the slow decay of the $p_z$ atomic-wavefunctions normal to the ring.    

	While the model presented thus far provides confidence that it is possible to identify the structures responsible for electron capture, the real challenge is to apply the insight gained to discriminate between isomers.  Toward this end, the isomer-dependent amplitude variations of a secondary electron resonance at about 1.0 eV are studied.  The energy of this resonance also varies little between isomers, again suggesting that the capturing orbital is $\pi$ in character.  For the isomers of nitrotoluene, we believe, electron capture into the LUMO is responsible for this resonance.  

	Shown in Figure 5 are the lowest unoccupied molecular orbitals, calculated using the ADF package, of 2-, 3-, and 4-nitrotoluene.  There are significant differences.  The local symmetry of 3-nitrotoluene and 4-nitrotoluene are approximately C$_{2v}$ by virtue of a near reflection symmetry in the plane normal to the ring containing the carbon to nitrogen bond.  However, 2-nitrotoluene has an asymmetric LUMO because of significant wave amplitude on the methyl group, reducing the local symmetry to C$_h$.  We believe this difference leads to the larger secondary resonance intensity of 2-nitrotoluene compared to the other two isomers (see Figure 1).

	Under the local C$_{2v}$ symmetry, the LUMOs of 3- and 4-nitrotoluene reduce as B$_1$.  Projecting this component from a general plane wave gives the wavefunction of the incident electron to be,
\begin{equation}
\phi(k_x, k_y, k_z) = i \cos(k_y y) \sin(k_z z) (\cos(k_x x) + i \sin(k_x x)) 
\end{equation}

As is evident, the reflection symmetry in the xz and yz-planes requires the center of a captured electronÕs wavefunction to sit on the x-axis.  There is only one set of wavevectors that simultaneously satisfy this constraint and connect regions of opposite phase.  The projection of these vectors in two planes is shown in Figure 6 along with the real component of the wavefunction of equation (8).  Though these displacement vectors are of a length expected for half the wavelength of 1.0 eV electron ( 12 \AA), the intervening region of opposite phase leads to significant cancellations between the wavefunctions of the incoming electron and the LUMO.  Thus, one can expect a weak resonance at about 1.0 eV, as is observed.  

	The symmetry reduction seen in the LUMO of the 2-nitrotoluene removes one of the constraints on the incident electron wavevector, confining its center only to the xy-plane.  No longer tethered to the x-axis, a new family of wavevectors (also   12 \AA in length) contributes to the reaction rate.  The projection of these wavevectors in the xy-plane is shown in Figure 7.  Because the center of these vectors has been displaced to a region of low LUMO amplitude, the cancellation will be reduced over those of 3- and 4-nitrotoluene, leading to a more pronounced 1.0 eV resonance, again, consistent with experimental observation.  

	While the model has shown itself to be consistent only with the ERS for the isomers of nitroroluene, we are confident that it is general, at least as far as the isomers of nitrated aromatics are concerned.  This optimism is based, in part, on the extensive molecular orbital calculations conducted on this series of compounds, most of which have not been reviewed here.    Still, the results are so promising that we believe a quantitative implementation of the model is in order.  Accordingly, we have begun the coding of equation (6) with the intent of predicting ERS for a great number of compounds.  The results will be discussed in a subsequent paper.

\section{Conclusions}
	We have shown that the utility of EM-MS as a tool for molecular identification can be enhanced through a two-step process.  In the first, chemical inference is employed to identify the molecular orbitals into which the monochromatic incident electrons are captured.  In the second step, the principles of quantum scattering theory can be employed to calculate the fragmentation rate, and thus the ion yield as a function of electron energies. Though for simple molecules, such as those studied here, such calculations are not necessary to explain spectral differences associated with different isomers.  For more complex molecules, the direct calculation of ion yield will be necessary.  The tools necessary to perform these calculations are presently under development.
We anticipate that these revelations will serve not only to accelerate the use of EM-MS for molecular identification, but also further stimulate a broader interest in the use of first principle methods with analytical chemistry.   

\section{ACKNOWLEDGMENTS.}  The authors thank Philip Morris USA and the Colorado Tobacco Research Program for support of this work.  The authors also extend thanks to John Dane and Giuseppe Petrucci for helpful discussions.

\vfill\eject

\vfill\eject

\begin{figure}
\includegraphics[width=2in]{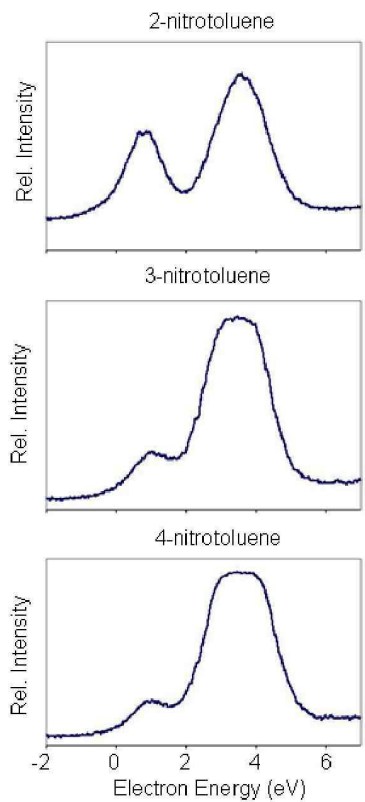}
\caption{\label{fig1} Ion yield curves for m/z 46 as a function of electron energy for the three isomers of mono-nitrotoluene obtained by electron monochromator-mass spectrometry.  }
\end{figure}

\begin{figure}
\includegraphics[width=3in]{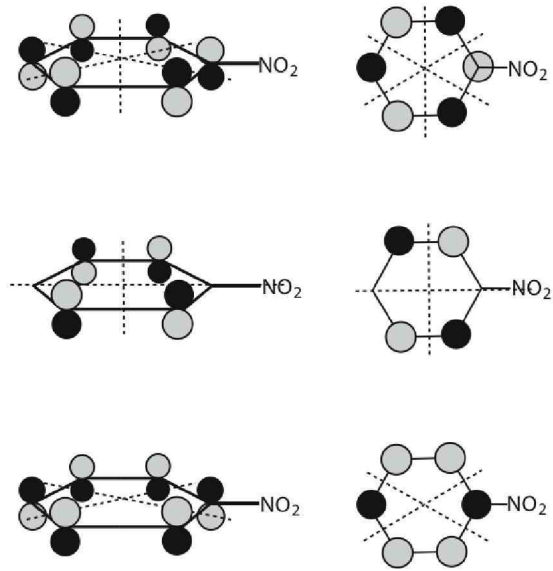}
\caption{\label{fig2} The unoccupied $\pi$-orbitals of a nitro aromatic compound, the orbital ``C'' is usually the LUMO while the ``B'' is typically the LUMO + 1.  }
\end{figure}

\begin{figure}
\includegraphics[width=3in]{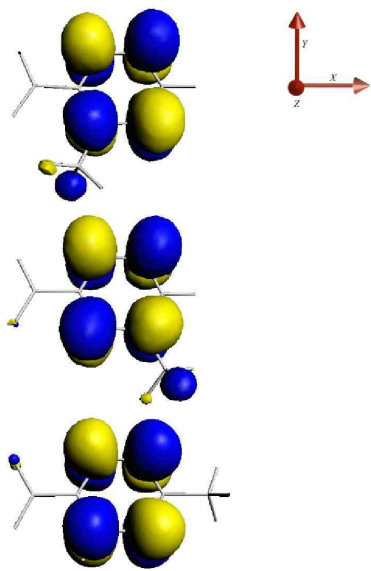}
\caption{\label{fig3} The LUMO + 1s of 2- (A), 3-(B), and 4-(C) nitrotoluene.  Note the similarity between these orbitals, all showing little amplitude on the nitro or methyl groups.  Thus the underlying symmetry can be taken as approximately D$_{2h}$. }
\end{figure}

\begin{figure}
\includegraphics[width=3in]{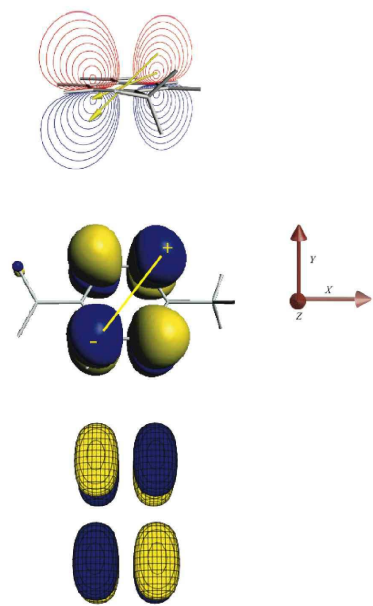}
\caption{\label{fig4}(A)  half-wavevectors that are ``in phase'' with the LUMO + 1 of 2-, 3-, and 4-nitrotoluene projected onto the plane normal to the ring and containing ring carbons 1 and 4 relative to the methyl group. In (B) another projection of this wavevector  onto the xy-plane of 4-nitrotoluene is shown for clarity.  The wavefunction corresponding to this wavevector is shown in (C).  Note the constructive overlap between the wavefunction and the LUMO. }
\end{figure}

\begin{figure}
\includegraphics[width=3in]{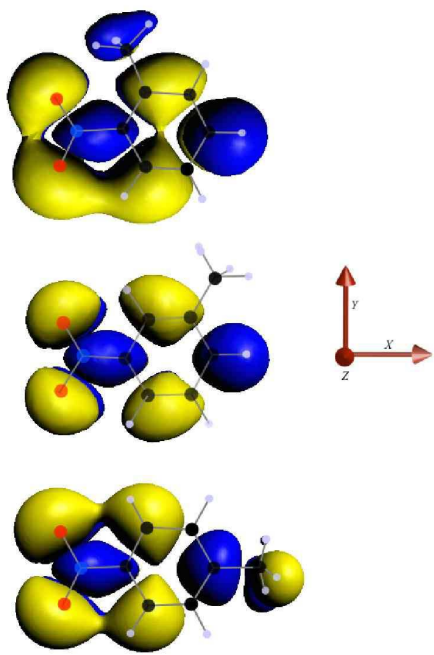}
\caption{\label{fig5}The LUMOs of 2-(A), 3-(B), and 4-(C) nitrotoluene.  Note the similarity between these orbitals B and C, which are of nearly C$_{2v}$ symmetry.  However, the significant amplitude on the methyl group of (A) 2-nitro toluene reduces its symmetry to C$_h$.  }
\end{figure}

\begin{figure}
\includegraphics[width=3in]{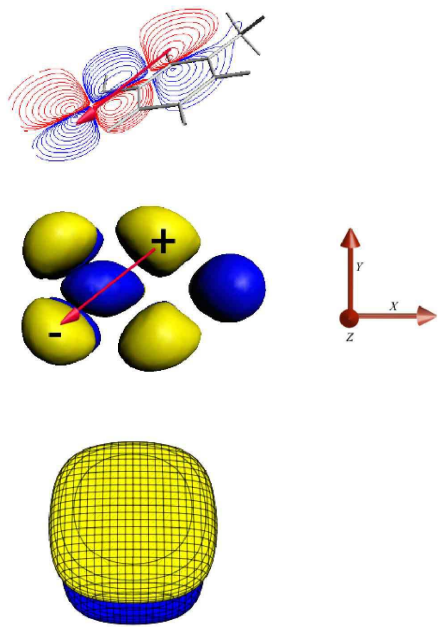}
\caption{\label{fig6}(A)  Half-wavevectors  that are ``in phase'' with the LUMOs of 2-, 3-, and 4-nitrotoluene projected onto the plane normal to the ring and containing a nitro oxygen and the carbon at position 2 relative to the nitro group. In (B) another projection of this wavevector onto the xy-plane of 4-nitrotoluene is shown for clarity.  The wavefunction corresponding to this wavevector is shown in (C).  Note that this wavefunction must be centered on the molecular x-axis for 3 and 4 nitrotoluene and so will experience both positive and negative overlap with the LUMO. }
\end{figure}

\begin{figure}
\includegraphics[width=3in]{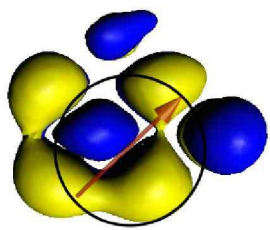}
\caption{\label{fig6} The local symmetry of 2-nitrotoluene is C$_h$ and the incident electron wavefunction must only be centered on the xy-plane.  At the position indicated, the negative overlap will be reduced, leading to a greater resonance signal. }
\end{figure}

\begin{table*}
\caption{\label{tab:table1} Resonance Energies for m/z 46 anions of nitro aromatic compounds (bond type: C-NO$_2$).}
\begin{ruledtabular}
\begin{tabular}{ccc}
 Compound&Resonance Energy (eV)&Normalized Relative Intensities (\%)\\ \hline
2-nitrotoluene&	0.8, 3.6&	60, 100 \\
3-nitrotoluene&	1.0, 3.5&	26, 100\\
4-nitrotoluene&	1.0, 3.6&	21, 100\\
2,6-dinitrotoluene&	0.6, 3.4&	21, 100\\
2-nitro-m-xylene&	1.0, 4.1&	100, 77\\
4-nitro-m-xylene&	1.0, 4.0&	39, 100\\
5-nitro-m-xylene&	1.1, 3.9&	14, 100\\
2-nitrophenol&	3.5&	100\\
4-nitrophenol&	1.3, 3.9&	14, 100\\
1-nitronaphthalene&	0.8, 3.1&	39, 100\\
2-nitronaphthalene&	1.0, 3.2&	16, 100\\
1,3-dinitronaphthalene&	3.4&	100\\
1,8-dinitronaphthalene&	0.7, 3.7&	40, 100\\
2-nitrobiphenyl&	1.0, 4.0&	100, 89\\
3-nitrobiphenyl&	1.1, 3.9&	23, 100\\
2,2Õ-dinitrobiphenyl&	0.6, 4.0&	64, 100\\
3,4Õ-dinitrobiphenyl&	1.0, 3.9&	31, 100\\
6-nitroquinoline&	1.2, 3.3&	24, 100\\
2-nitroanthracene&	2.7&	100\\
9-nitroanthracene&	1.4, 4.0&	100, 96\\
9,10-dinitroanthracene&	4.7&	100\\
4-nitrophenanthrene&	0.7, 3.9&	57, 100\\
1-nitropyrene&	3.6&	100\\
1,3-dinitropyrene&	4.0&	100\\
3-nitrofluoranthene&	2.8, 4.7&	100, 77\\
\end{tabular}
\end{ruledtabular}
\end{table*}

\end{document}